\documentclass[a4,12pt]{article}
\usepackage{amssymb}
\usepackage[dvips]{graphicx,color}
\usepackage{epsfig}

\begin{document}

{\bf\Large
\begin{center}
On the Existence of a Self-Similar Coarse Graining of a Self-Similar Space
\end{center}
}

\begin{center}
Akihiko Kitada, Tomoyuki Yamamoto, Tsuyoshi Yoshioka and Shousuke Ohmori 

Laboratory of mathematical design for materials,\\ Faculty of Science and Engineering, Waseda University,\\ 3-4-1 Okubo, Shinjuku-ku, Tokyo 169-8555
\end{center}

\noindent{\bf Abstract}\\
A topological space homeomorphic to a self-similar space is demonstrated to be self-similar. There exists a self-similar space $S$ whose coarse graining is homeomorphic to $S$. The coarse graining of $S$ is, therefore, self-similar again. In the same way, the coarse graining of the self-similar coarse graining of $S$ is, furthermore, self-similar. These situations succeed endlessly. Such a self-similar $S$ is generated actually from an intense quadratic dynamics.

\bigskip


\section{Introduction}

In the fractal sciences, the fine structure of the self-similar space is characterized by the property that every details looks similar with the whole. In the present report, we are oppositely concerned with the coarse structures of a self-similar space, that is, with the problem "what self-similar space can have a coarse graining of it with a self-similarity again?". 
According to A. Fern$\rm\acute{a}$ndez \cite{ref1}, the procedure of the coarse graining or the block construction \cite{ref2} of a space in the statistical physics corresponds mathematically to that of the construction of a quotient space which is defined by a classification of all points in the space through the identification of the different points based on an equivalence relation.

At first, a sufficient condition for a given topological space to be metrizable and self-similar with respect to the metric is investigated, and, second, the existence of a decomposition space \cite{ref3} as a coarse graining of a self-similar space $S$ whose self-similarity is defined by a system of weak contractions which is topologically closely related to that defining the self-similarity of $S$ is discussed in a quite elementary way.
As a consequence, we are convinced that there exists a sequence of self-similar coarse graining of a self-similar space even for the quadratic dynamics known to be one of the simplest dynamical system.
Finally, it is noted that each step of the sequence can equally generate a topological space characteristic of condensed matter such as dendrite \cite{ref4}.

\section{A condition for a topological space to be self-similar}

An answer of the problem "for what topological space, can we find a system of weak contractions which makes the space self-similar?" is simply stated as follows.

\bigskip

{\noindent\bf Proposition.}~{\it The existence of a self-similar space which is homeomorphic \cite{ref5} to $(Y,\tau)$ is sufficient for a topological space $(Y,\tau)$ to be a metrizable space and self-similar with respect to the metric.}

{\noindent\bf Proof.}~~{ Let $(X,\tau_d)$ be self-similar based on a system of weak contractions \\ $p_j:(X,\tau_d)\to(X,\tau_d)$, $d(p_j(x),p_j(x'))\le\alpha_j(\eta)d(x,x')$ for $d(x,x')<\eta,~0\le\alpha_j(\eta)<1,~j=1,\ldots,m~(2\le m<\infty)$. That is, $\displaystyle\bigcup_{j=1}^mp_j(X)=X$. Using a homeomorphism $h:(X,\tau_d)\simeq(Y,\tau)$, we can define a metric $\rho$ on $Y$ as 
$$
\rho(y,y')=d(h^{-1}(y),h^{-1}(y')),~~y,~y'\in Y.
$$
The metric topology $\tau_\rho$ is identical with the initial topology $\tau$ . From the relations 1) and 2) below, the metric space $(Y,\tau_\rho)$ is confirmed to be self-similar by a system of weak contractions $q_j:(Y,\tau_\rho)\to(Y,\tau_\rho),~j=1,\ldots,m$ where $q_j$ is topologically conjugate to $p_j$ with the above homeomorphism $h$, that is, $q_j=h\circ p_j\circ h^{-1}$ .
\begin{itemize}
\item[1)]$\rho(q_j(y),q_j(y'))=d(h^{-1}(q_j(y)),h^{-1}(q_j(y')))$\\
$=d(p_j(h^{-1}(y)),p_j(h^{-1}(y')))\le\alpha_j(\eta)d(h^{-1}(y),h^{-1}(y'))$\\
$=\alpha_j(\eta)\rho(y,y')$ for $\rho(y,y')<\eta$.
\item[2)]$\displaystyle\bigcup_{j=1}^mq_j(Y)=\bigcup_{j=1}^mq_j(h(X))=h(\bigcup_{j=1}^mp_j(X))=h(X)=Y.$  ~~$\Box $
\end{itemize}

\section{Existence of a self-similar decomposition space}

As an application of Proposition, we will show the existence of a self-similar decomposition space of a self-similar space.

Let $S$ be a self-similar, perfect \cite{perfect}, zero-dimensional (0-dim) \cite{ref6}, compact metric space,  and $(X,\tau_d)$ be any compact metric space which is self-similar.
 Then, there exists a continuous map $f$ from $S$ onto $X$ \cite{ref7}, and $X$ is homeomorphic to the decomposition space $(\mathcal{D}_f,\tau(\mathcal{D}_f))$ of $S$ with a homeomorphism $h:(X,\tau_d)\simeq(\mathcal{D}_f,\tau(\mathcal{D}_f)),~x\mapsto f^{-1}(x)$ \cite{ref8}. 
 Here, $\mathcal{D}_f=\{f^{-1}(x)\subset S;~x\in X\}$ and $\displaystyle\tau(\mathcal{D}_f)=\{\mathcal{U}\subset\mathcal{D}_f;~\bigcup\mathcal{U}(=\bigcup_{D\in\mathcal{U}}D)$ is an open set of $S\}$. 
 The decomposition topology $\tau(\mathcal{D}_f)$ is identical with a metric topology $\tau_\rho$ with a metric $\rho(y,y')=d(h^{-1}(y),h^{-1}(y')),y,~y'\in\mathcal{D}_f$ \cite{ref9}. 
 Since the metric space $(X,\tau_d)$ is assumed to be self-similar, from Proposition, the decomposition space $(\mathcal{D}_f,\tau_\rho)$ must be self-similar based on a system of weak contractions each of which is topologically conjugate to each weak contraction which defines the self-similarity of $X$.
According to the self-similarity of the selected space $X$, the decomposition space $\mathcal{D}_f$ of $S$ can have various types of self-similarity.

Now, let us consider a special case where the system of contructions defining the self-similarity of the decomposition space $\mathcal{D}_f$ of $S$ is topologically related to that defining the self-similarity of $S$. Let $\{S_1, \cdots, S_n\}$ be a partition of $S$ \cite{ref3} such that each $S_i$ is a clopen (closed and open) set of $S$. (Concerning the existence of such partition of $S$, see Appendix.) 
Since the metric space $S_1$ is perfect, 0-dim, compact, it is homeomorphic to the Cantor's Middle Third Set (abbreviated to CMTS) \cite{ref10} as well as the space $S$. 
Therefore, $S_1$ and $S$ are homeomorphic.
Let $f: S \to S_1$ be a \it{not one to one}\rm, continuous, onto map.  
For example, the map $f:S \to S_1$ defined as $f(x)=x$ for $x \in S_1, f(x) \equiv q_2 \in S_1$ for $x \in S_2, \cdots, f(x) \equiv q_n \in S_1$ for $x \in S_n$ is a continuous, onto map. 
It must be noted that $\mathcal{D}_f$ is not trivial decomposition space $\{\{x\} \subset S; x \in S\}$ because the map $f$ is not one to one \cite{add12}. 
Since the decomposition space $\mathcal{D}_f$ of $S$ is homeomorphic to $S_1$ \cite{ref8}, $S$ must be homeomorphic to $\mathcal{D}_f$. 
Therefore, from Proposition, $\mathcal{D}_f$ is self-similar based on a system of weak contractions each of which  is topologically conjugate to each weak contraction which defines the self-similarity of $S$. 

Since the metric space $\mathcal{D}_f$ is perfect, 0-dim and compact, the same situation as for the initial space $S$ can take place for the decomposition space $\mathcal{D}_f$ of $S$. 
Therefore, continuing this process endlessly, we obtain an infinite sequence of self-similar decomposition spaces or self-similar coarse graining starting from the self-similar space $S$, namely, a hierarchic structure of self-similar spaces as shown in Fig.1. In Fig. 1, the above mentioned decomposition space $\mathcal{D}_f$ of $S$ is denoted by $\mathcal{D}^1$. $\mathcal{D}^1$ is self-similar due to a system of weak contractions $\{f^1_j =h^1 \circ f_j \circ (h^1)^{-1}: \mathcal{D}^1 \to \mathcal{D}^1; j=1, \cdots, m\}$. Here, $\{f_j: S \to S; j=1, \cdots, m\}$ is a system of weak contractions which defines the self-similarity of $S$, and $h^1$ is a homeomorphism from $S$ to $\mathcal{D}^1$. The decomposition space $\mathcal{D}^2$ of $\mathcal{D}^1$ is self-similar based on a system of weak contractions $\{f^2_j=h^2 \circ f^1_j \circ (h^2)^{-1}:\mathcal{D}^2 \to \mathcal{D}^2; j=1, \cdots, m \}$ where $h^2$ is a homeomorphism from $\mathcal{D}^1$ to $\mathcal{D}^2$. We can continue the procedure in this manner.

\bigskip

{\noindent\bf Statement.} \cite{ref12,ref13, ref14} ~{\it Let $(Z,\tau_d)$ be a compact metric space. If the sytem $\{f_j:(Z,\tau_d)\to(Z,\tau_d),~j=1,\ldots,m\}$ of weak contractions \\
$d(f_j(z),f_j(z'))\le\alpha_j(\eta)d(z,z')~{\rm for}~d(z,z')<\eta,~0<\alpha_j(\eta)<1,$
$\inf_{\eta>0}\alpha_j(\eta)>0, j= 1, \cdots, m$ \\
satisfies three conditions
\begin{itemize}
\item[i)] Each $f_j$ is one to one,
\item[ii)] The set $\bigcup_{j=1}^m\{z\in Z;~f_j(z)=z\}$ is not a singleton,
\item[iii)] $\sum_{j=1}^m inf_{\eta>0} \alpha_j(\eta)<1$,
\end{itemize}
then, there exists a perfect, 0-dim, compact $S~(\subset Z)$ such that \\ $\bigcup_{j=1}^mf_j(S)=S$.}

\bigskip

Concludingly, we are convinced of the existence of a sequence as shown in Fig. 1 of self-similar coarse graining of a self-similar space based on the above quadratic dynamics $F_{\mu}(x)$ with a sufficiently large rate constant $\mu > 0$.

\section{Generation of dendrites from each step of the sequence $S, \mathcal{D}^1, \mathcal{D}^2, \cdots$}

Since all of the metric spaces $S, \mathcal{D}^1, \mathcal{D}^2, \cdots$ in Fig. 1 are perfect, $0$-dim and compact, there exist continuous maps \cite{ref7}, $k$ from $S$ onto the dendrite $\delta$ as a compact metric space, $k^1$ from $\mathcal{D}^1$ onto $\delta$, $k^2$ from $\mathcal{D}^2$ onto $\delta, \cdots$, respectively \cite{ref15}. 
The decomposition spaces $\delta_S=\{k^{-1}(x) \subset S ; x \in \delta\}$  of $S$ due to $f$, $\delta_{\mathcal{D}^1}=\{(k^1)^{-1}(x) \subset \mathcal{D}^1 ; x \in \delta\}$ of $\mathcal{D}^1$ due to $k^1$, $\delta_{\mathcal{D}^2}=\{(k^2)^{-1}(x) \subset \mathcal{D}^2 ; x \in \delta\}$ of $\mathcal{D}^2$ due to $k^2$, $\cdots$ are homeomorphic to the dendrite $\delta$, and therefore, $\delta_S, \delta_{\mathcal{D}^1}, \delta_{\mathcal{D}^2}, \cdots$ must have the dendritic structure in common (Fig. 3). 
For example, the self-similar space $S$ generated from a quadratic dynamics $F_{\mu}(x) = \mu x(1-x)$ with a sufficiently large $\mu > 0$ is mathematically demonstrated to be able to form a dendrite through the coalescence or the rearrangement of constituents of $S$.

\bigskip
{\noindent\bf Appendix}

Let $S$ be a perfect, 0-dim $T_0$-space. Then, for any $n$, there exist $n$ non-empty clopen (closed and open) sets $S_1, \cdots, S_n$ of $S$ such that $S_i \cap S_{i'}=\phi$ for $i \ne i'$ and $\displaystyle\bigcup_{i=1}^n S_i =S$. For any $n$, there exist $n$ non-empty clopen sets $S_{i_1}, \cdots, S_{i_n}$ of $S$ such that $S_{i_j} \cap S_{i_{j'}} = \phi$ for $j \ne j'$ and $\displaystyle\bigcup_{j=1}^n S_{i_j}=S_i$. We can continue in this manner endlessly.

{\noindent\bf proof)} To use the mathematical induction, let the statement hold for $n-1$. Since $S$ is perfect, the open set $S_{n-1}$ has at least two distinct points $a$ and $b$. Since $S$ is a $T_0$-space, there exists an open set $u$ containing $a$ such that $b \notin u$ without loss of generality. Since $S$ is 0-dim, there exists a clopen set $v$ which contains the point $a$ and is contained in the open set $u \cap S_{n-1}$. Since $b \in S_{n-1}-v$, the clopen set $S_{n-1}-v$ is not empty. Thus, we obtain a desired $n$-partition $\{S_1, \cdots, S_{n-2}, v, S_{n-1}-v\}$ of $S$. Concerning the subspace $S_i$, it suffices to remember that any non-empty open set in a perfect space is perfect again. 
$\Box$

\bigskip

{\noindent\bf Acknowledgment}

The authors are grateful to Professor H. Fukaishi of Kagawa University for useful discussions.

\newpage

\begin{figure}
\begin{center}
\includegraphics[width=12cm]{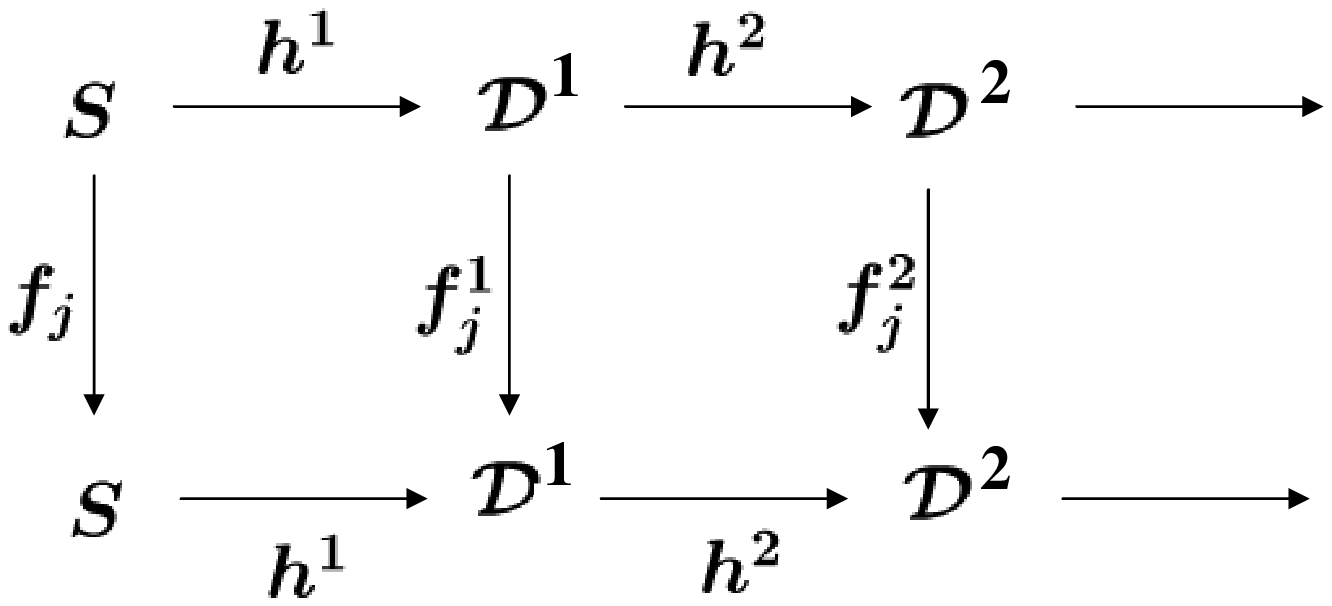}
\end{center}
\caption{A hierarchic structure of self-similar spaces. $h^i, i=1, 2, \cdots$ are homeomorphisms. $f_j, f_j^1, f_j^2, \cdots$ are weak contractions such that $\displaystyle\bigcup^m_{j=1} f_j(S)=S, \displaystyle\bigcup^m_{j=1} f_j^1(\mathcal{D}^1)=\mathcal{D}^1, \bigcup^m_{j=1} f_j^2(\mathcal{D}^2)=\mathcal{D}^2, \cdots$, respectively.}
\label{f1}
\end{figure}

~~~
\bigskip

\begin{figure}
\begin{center}
\includegraphics[width=12cm]{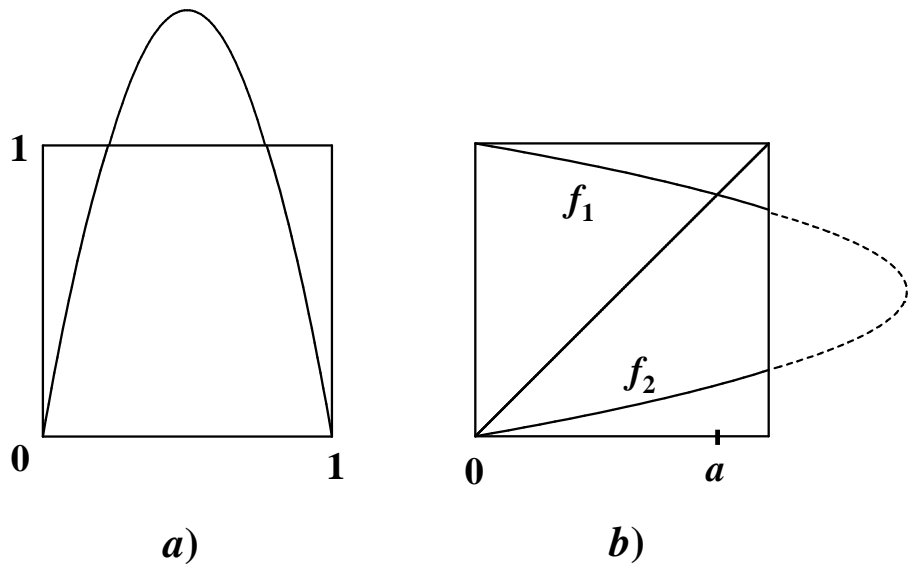}
\end{center}
\caption{a) $F_{\mu}(x) = \mu x (1-x), ~~\mu > 4, ~~x \in [0, 1]$. b) The quadratic dynamics $F_{\mu}(x)$  defines a system of contractions $\{f_j:[0, 1] \to [0, 1], j=1, 2 \}$ which satisfies three conditions \it{i), ii), iii)} \rm  in \bf{Statement} \rm  in the text. In fact, $\displaystyle \bigcup_{j=1,2} \{x \in [0,1]; f_j(x)=x\}=\{0,a\}$.}
\label{f1}
\end{figure}

~~~
\newpage

\begin{figure}
\begin{center}
\includegraphics[width=12cm]{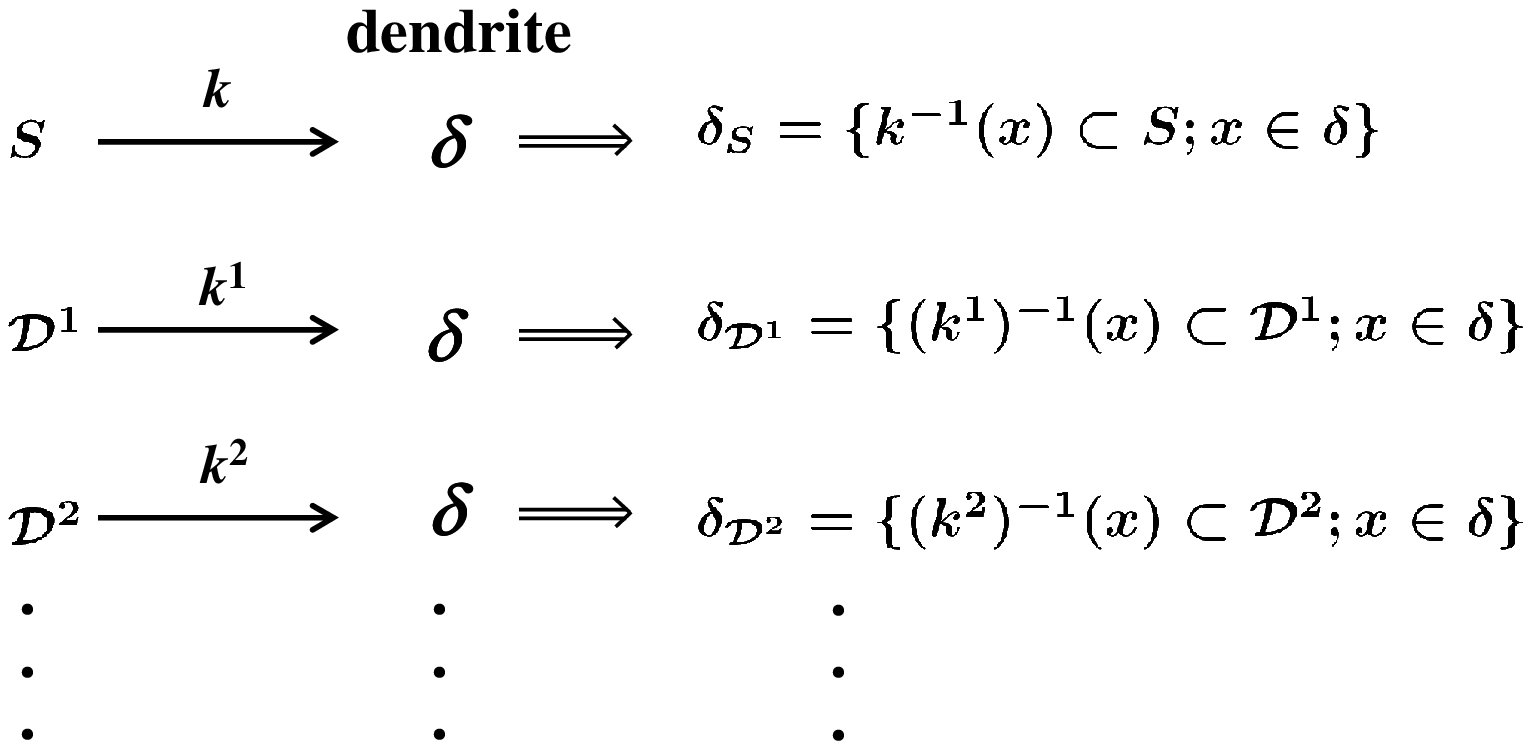}
\end{center}
\caption{Generation of dendrites from each step of the sequence $S, \mathcal{D}^1, \mathcal{D}^2, \cdots$.  $\delta, \delta_S, \delta_{\mathcal{D}^1}, \delta_{\mathcal{D}^2}, \cdots$ are dendrites. }
\label{f1}
\end{figure}

\newpage


\begin{thebibliography}{9}
\bibitem{ref4} A dendrite is a metric space which is locally connected, connected and compact. The reference in [3], p.165.
\bibitem{perfect} A topological space $(A, \tau)$ is said to be perfect provided that any set $\{x\}$ composed of single point $x \in A$ is not an open set, that is, $\{x\} \notin \tau$.
\bibitem{ref8} The reference in [3], p.44. The decomposition space $\mathcal{D}_f$ of $S$ can have various types of topological structure. For example, if $X$ is a dendrite (the reference in [3], p.165), also $\mathcal{D}_f$ must be dendrite.
\bibitem{add12} One of the simplest example of such decomposition $\mathcal{D}_f$ is as follows. \\ Let the self-similar, perfect, 0-dim, compact metric space $S$ be CMTS itself and let the partition $\{S_1, S_2\}$ of $S$ be the set $\{$CMTS $\cap [0,1/3],$ CMTS $\cap [2/3,1]\}$. Then, the set $\{f^{-1}(x) \subset$ CMTS $, x \in S_1\}$ where $f^{-1}(x) =\{x\} \subset X$ for $ x \in S_1 - \{q\}$ and $f^{-1}(q) =\{q\} \cup S_2$, is a decomposition $\mathcal{D}_f$ of CMTS.
\bibitem{ref15} It is noted that continuous maps $k, k^1, k^2, \cdots$ must not be one to one. In fact, if they are one to one, they must be homeomorphisms between 0-dim (i.e., totally disconnected) spaces $S, \mathcal{D}^1, \mathcal{D}^2, \cdots$ and a connetcted space $\delta$. It is impossible.
\end{thebibliography}
\end{document}